\documentstyle[aps,prl]{revtex}
\begin{document}
\draft
\twocolumn[\hsize\textwidth\columnwidth\hsize\csname @twocolumnfalse\endcsname
\title{Novel approach to  description of  spin liquid phases 
in low-dimensional quantum antiferromagnets.}
\author{V.N. Kotov, O. Sushkov, Zheng Weihong, and J. Oitmaa}
\address{School of Physics, University of New South Wales, Sydney 2052, Australia}

\maketitle

\begin{abstract}
We consider quantum spin systems with dimerization,  which
at strong coupling  have singlet ground states.
 To account for strong correlations, the $S=1$ elementary excitations
are described as dilute Bose gas with infinite on-site repulsion. 
This approach is applied to 
the   two-layer Heisenberg antiferromagnet  at  $T=0$  with general couplings.
Our analytic  results for the triplet gap, the excitation
spectrum and the location of the quantum critical point  are in
excellent agreement with numerical results, obtained by   dimer series expansions. 
\end{abstract}

]


One of the most challenging problems of quantum magnetism
is the description of transitions between phases with spontaneously
broken symmetry and disordered (spin liquid) phases. 
The properties of the disordered phases are also of great interest
\cite{Subir}.
 
A variety of quantum spin models  have been 
introduced in connection with the high-$T_{c}$ cuprates and other
recently discovered  compounds.
 Examples include the Heisenberg ladder \cite{Copalan}, 
 the two-layer Heisenberg model \cite{ChMorr} and 2D square lattice models with
dimerization \cite{Kato}.
 In all of the above the Hamiltonian
favors singlet formation of the spins between the chains (layers) or
on neighboring sites.
For this class of models the  disordered phase is relatively well understood, 
since the lowest excitation above the singlet is a massive
triplet.
Another example is the  $CaV_{4}O_{9}$ lattice
\cite{CAVO}, where the spins  form a singlet state on a plaquette.
There also have been suggestions that dimerization of different kinds may
occur in the  $J_{1}-J_{2}$ model   
\cite{Sachdev,Zhitomirsky}.

All of the models mentioned above, except for the ladder,
exhibit a quantum phase transition from a disordered dimer phase
to a collinear N\'{e}el phase with long range order in the ground state 
as the dimerization decreases. This transition occurs due to competition
between singlet formation and antiferromagnetic order.
A useful approach to  the description of the disordered  phase
 is the bond operator representation for
spins, introduced by Chubukov \cite{Chubukov} and Sachdev and Bhatt
 \cite{Sachdev}. This representation  can be considered as  
the analog  
of the usual Holstein-Primakoff transformation for  phases with unbroken
spin rotational symmetry. Let us consider two $S=1/2$ spins $\vec{S}_{1}$,
 $\vec{S}_{2}$ and introduce operators for creation 
of a singlet $s^{\dagger}|0> = \frac{1}{\sqrt{2}}(|\uparrow
\downarrow> - |\downarrow \uparrow>)$ 
and    three triplet states $t_{\alpha}^{\dagger}, 
\alpha = x,y,z$  above a fictitious vaccum $|0>$:
$t_{x}^{\dagger}|0> =-\frac{1}{\sqrt{2}} (|\uparrow \uparrow> - |\downarrow \downarrow
>), t_{y}^{\dagger}|0> = i \frac{1}{\sqrt{2}}(|\uparrow \uparrow> + 
|\downarrow \downarrow>), t_{z}^{\dagger}|0> =  \frac{1}{\sqrt{2}}(|\uparrow 
\downarrow> + |\downarrow \uparrow>)$. 
 Then the following representation is exact\cite{Sachdev}:

\begin{equation}
S_{1,2}^{\alpha} = \frac{1}{2} ( \pm s^{\dagger}  t_{\alpha} \pm
t_{\alpha}^{\dagger} s  - i \epsilon_{\alpha\beta\gamma} t_{\beta}^{\dagger} 
t_{\gamma})
\end{equation}
The four operators satisfy the usual bosonic commutation relations.
In order to ensure that the physical states are either singlets or
triplets one has to impose the condition:
$s^{\dagger}s + t_{\alpha}^{\dagger}t_{\alpha} = 1$. 
For a lattice spin system, the constraint is typically   taken
into account in  a mean-field fashion, i.e. it is not strictly satisfied
on every site, but only on average \cite{Sachdev}. 
A slightly different representation can be obtained by
choosing the singlet as the ground state. Then Eq.(1) is still valid,
but the operator $s$ has the form: $s = \sqrt{1-t_{\alpha}^{\dagger}t_{\alpha}}$,
which formally is the resolution of the constraint\cite{Chubukov,ChMorr}.
Again, the form of $s$ ensures that only physical states are present.
However, it is  very difficult to take the $s$  term into account
due to its non-linear nature. Expansions of the square
root to infinite order have been proposed \cite{ChMorr}. Unfortunately,
there is no small parameter in this expansion and therefore
the summation is ambiguous and  technically complicated.
 Alternatively, one can   use numerical techniques, based on the Gutzwiller 
projection method  \cite{Eder2}.


In  this Letter we present an effective analytical method to deal
with the hard core constraint. This approach  can be applied to any
model, for which the excitations in the disordered phase are
triplets above a strong coupling singlet ground state. For
definiteness we consider the model:  

\begin{equation}
H = J \sum_{<i,j>} \vec{S}_{1i}.\vec{S}_{1j} +
\lambda J \sum_{<i,j>} \vec{S}_{2i}.\vec{S}_{2j}+
J_{\perp}\sum_{i} \vec{S}_{1i}.\vec{S}_{2i}.
\end{equation}
All the spins are $1/2$ and the couplings are antiferromagnetic
($J, J_{\perp} \geq 0$). The spins $\vec{S}_{1i}, \vec{S}_{2i}$
represent two planes of Heisenberg spins, coupled through the third
term in (2). The summation in each plane is over nearest neighbors
on a square lattice.
In the present work we consider two cases: 
$\lambda = 1$, which corresponds to the two-layer Heisenberg model, and $\lambda = 0$,
describing   free spins in one of the planes. The latter model is interesting
because of its connection to the Kondo  lattice model (at half filling)
with an additional repulsive Hubbard interaction between the conduction
electrons. In the limit when the repulsion  is strong,
the charge degrees of freedom are frozen, while the spin part
is described by the  Heisenberg Hamiltonian, leading to (2) (at $\lambda =0$). 
A simplified version of  this model was introduced by Doniach \cite{Doniach}
to study the competition between  local singlet formation (the Kondo effect)
and the induced magnetic (RKKY) interaction  between the free spins.
The early mean-field treatment of Doniach predicted a critical $J_{\perp}$, below which
the free spins order antiferromagnetically.
Recently this model was also studied numerically by Matsushita, Gelfand and
Ishii \cite{Gelfand} who also found a finite transition point. 

For $J_{\perp} >> J$ interplane singlets are favored and the wave function
is a product of on-site dimers. The excitations above this strong coupling
ground state are triplets. In order to obtain the effective Hamiltonian
for the triplets we pair the spins into inter-plane singlets by using (1).
Alternatively, instead of applying the transformation (1), one could
use perturbation theory in the "hopping" $J$, and calculate
matrix elements of the type: $<t_{\alpha i},s_{j}|\vec{S}_{1i}.\vec{S}_{1j}
|s_{i},t_{\alpha j}> = <t_{\alpha i},t_{\alpha j}|
\vec{S}_{1i}.\vec{S}_{1j}|s_{i},s_{j}>
=1/4, <t_{\alpha i},t_{\beta j}|\vec{S}_{1i}.\vec{S}_{1j}|t_{\gamma i},
t_{\delta j}> = 1/4(
\delta_{\alpha \delta}\delta_{\gamma \beta} -
\delta_{\alpha\beta}\delta_{\gamma\delta})$, etc.
The latter method is more useful when additional degrees  of
freedom are present in the problem, e.g. holes.
For a start we neglect the constraint completely (i.e. formally
set $s=1$ in (1)) and obtain  the effective Hamiltonian:

\begin{equation}
H = H_{2} +  H_{3}  +  H_{4},
\end{equation}
\begin{equation}
H_{2} = \sum_{\bf{k}, \alpha} A_{\bf{k}} t_{\bf{k}\alpha}^{\dagger}t_{\bf{k}\alpha}
+ \frac{B_{\bf{k}}}{2}\left(t_{\bf{k}\alpha}^{\dagger} 
 t_{\bf{-k}\alpha}^{\dagger}
+ \mbox{h.c.}\right)
\end{equation}


\begin{equation}
H_{3} = \frac{(\lambda-1)J}{4} \sum_{<i,j>,\alpha\beta\gamma}
\left\{[i\epsilon_{\alpha\beta\gamma}  t_{\alpha i}^{\dagger}
t_{\beta j}^{\dagger} t_{\gamma j} + \mbox{h.c.} ] + [i 
\leftrightarrow j] \right\}
\end{equation}
     

\begin{equation}
H_{4} = \frac{(1+\lambda)J}{4} \sum_{<i,j>,\alpha\beta} 
\left\{ t_{\alpha i}^{\dagger}t_{\beta j}^{\dagger}t_{\beta i}
t_{\alpha j} - t_{\alpha i}^{\dagger}t_{\alpha j}^{\dagger} 
t_{\beta i}t_{\beta j} \right\}
\end{equation}
The coefficients in (4) are: 
$A_{\bf{k}} = J_{\perp} + (1+\lambda)J\xi_{\bf{k}}$,
$B_{\bf{k}} = (1+\lambda)J\xi_{\bf{k}}$,
where $\xi_{\bf{k}} = (\mbox{cos}(k_{x}) + \mbox{cos}(k_{y}))/2$.
By using the Bogoliubov transformation 
$ t_{{\bf{k}} \alpha} = u_{\bf{k}} \tilde{t}_{{\bf{k}} \alpha} 
+  v_{\bf{k}} \tilde{t}_{-{\bf{k}} \alpha}^{\dagger}$
 we obtain for the excitation spectrum
at the quadratic level ($H_{2}$ only):
$\omega_{\bf{k}}^{2} = A_{\bf{k}}^{2} - B_{\bf{k}}^{2}$.
The gap $\Delta = \omega_{\pi,\pi}$ is non-zero for
$J_{\perp} > (J_{\perp})_{c} = 2(1+\lambda)J$ and vanishes at
$(J_{\perp})_{c}$, signaling a transition to a N\'{e}el ordered phase.
The location of the critical point at this level of approximation
$(J_{\perp})_{c} = 2J \ (\lambda =0),\ 4J \ (\lambda =1)$  differs
significantly from  the recent numerical results
1.39 (\cite{Gelfand}, this work) and 2.54 \cite{Zheng}, respectively.
Let us mention that  spin wave theory in the ordered phase works rather
poorly for this problem, predicting  for $\lambda =1$, 
$(J_{\perp})_{c} \approx 4.3J$ \cite{ChMorr}.

 We find, in agreement with previous work \cite{Sachdev,Copalan},  
 that  the effect  of the terms $H_{3}$ and
$H_{4}$ on the spectrum is quite small  
 and therefore  can not explain the numerical results.
We  treat these terms later  perturbatively.

The dominant contribution to the renormalization of the spectrum
comes from   the constraint  that
only one of the triplet states can be excited on every site: 
$t_{\alpha i}^{\dagger}t_{\beta i}^{\dagger} =0$.
This hard-core condition can be taken into account  by introducing an infinite
on-site repulsion between the bosons:


\begin{equation} H_{U} = U \sum_{i,\alpha \beta}
t_{\alpha i}^{\dagger}t_{\beta i}^{\dagger}t_{\beta i}t_{\alpha i},
 \ \ U \rightarrow \infty
\end{equation}
Since the interaction is infinite, one has to find the exact 
scattering amplitude  for the triplets. Our treatment is similar
to  the one used  for  Fermi gas with hard core,
which appears in the theory of nuclear matter and $^{3}He$.
The approach was initiated by Brueckner \cite{Brueckner}.  
The scattering vertex $\Gamma_{\alpha\beta,\gamma\delta} ({\bf{K}}),
 \ {\bf{K}} \equiv ({\bf{k}}, \omega)$ 
 in the ladder
approximation satisfies the Bethe-Salpeter equation, shown
in Fig.1a. It depends on the total energy and momentum of the incoming
particles ${\bf{K}} = {\bf{K}}_{1} + {\bf{K}}_{2}$ and has the structure
$\Gamma_{\alpha\beta,\gamma\delta} = \Gamma(\delta_{\alpha\gamma}
\delta_{\beta\delta} + \delta_{\alpha\delta}\delta_{\beta\gamma})$.
 Since the interaction is local and non-retarded,
the equation for $\Gamma$ can be readily solved with the result:

\begin{eqnarray}
\Gamma({\bf{K}}) & = & i\left( \int \frac{ d^{3}Q}{(2\pi)^3}
G({\bf{Q}})G({\bf{K}} - {\bf{Q}}) \right)^{-1} \nonumber \\
=&-&\left(\frac{1}{N} \sum_{\bf{q}} \frac{u_{\bf{q}}^{2}
u_{\bf{k}- \bf{q}}^{2}}{\omega - \omega_{\bf{q}} - \omega_{\bf{k}- \bf{q}}}
 + \left\{ \begin{array}{c} u \rightarrow v \\
\omega \rightarrow -\omega \end{array} \right\}\right)^{-1}
\end{eqnarray}
Here $G({\bf{Q}})$ is the normal triplet Green's function,
i.e. $G({\bf{k}},t)=-i<T(t_{\bf{k} \alpha}(t)t_{\bf{k} \alpha}^{\dagger}(0))>$
and the Bogoliubov coefficients  $u_{\bf{k}}^{2}, v_{\bf{k}}^{2} =
\pm 1/2 + A_{\bf{k}}/2\omega_{\bf{k}}$. 
The imaginary part of $\Gamma$ is determined by the rule
$\omega \rightarrow \omega + i\delta$.  

The basic approximation made  in the derivation of $\Gamma({\bf{K}})$
is that we neglect all anomalous scattering vertices, which are
present in the  theory due to  the existence of anomalous Green's functions,
$G_{a}({\bf{k}},t)=-i<T(t_{-\bf{k} \alpha}^{\dagger}(t)
t_{\bf{k} \alpha}^{\dagger}(0))>$.
We have also  derived the complete set of equations by 
taking all vertices into account. However, our key observation is that all  anomalous
contributions are suppressed by an additional small parameter, present
in the theory - the density of triplet excitations 
$n_{i} = 
\sum_{\alpha}<t_{\alpha i}^{\dagger}t_{\alpha i}> =
3 N^{-1}\sum_{\bf{q}}v_{\bf{q}}^{2}
 \approx 0.1$ at $J_{\perp}/J \approx 2.5$. We find that $n_{i}$
is quite small  throughout the disordered phase, even close to the
transition point. Thus the triplet excitations behave as a dilute, 
strongly interacting Bose gas. Consequently, since 
an   insertion of an anomalous Green's function into the
intermediate states of the ladder in Fig.1a  brings powers 
of $v_{\bf{q}}$ into the equation for the
amplitude, its contribution is small. Therefore Eq.(8) can be considered as
the  first term in an expansion in powers of  the gas parameter $n_{i}$.
To be consistent, we also neglect the second term in (8), since
it contains $v_{\bf{q}}$.

The self-energy, corresponding to the scattering amplitude
  $\Gamma$ is found as a sum
of the diagrams shown in Fig.1b:
\begin{equation}
\Sigma({\bf{k}},\omega) = \frac{4}{N}\sum_{\bf{q}} v_{\bf{q}}^{2}
\Gamma({\bf{k}} + \bf{q}, \omega - \omega_{\bf{q}}). 
\end{equation}
Let us stress again   that  at our level of approximation (dilute gas), there  
is only a normal self-energy.
Next, in order to find the renomalized spectrum, one has to solve
the coupled Dyson equations for the normal and anomalous Green's
functions. Since the procedure is well know from the theory of a Bose
gas, we only write the final result for the normal Green's function 
\cite{Abrikosov}:

\begin{equation}
G({\bf{k}},\omega) = \frac{\omega + A_{\bf{k}} + \Sigma({\bf{k}},-\omega)}
{[\omega + A_{\bf{k}} + \Sigma({\bf{k}},-\omega)][\omega -  A_{\bf{k}}
- \Sigma({\bf{k}},\omega)]+ B_{\bf{k}}^{2}}
\end{equation}
After separating this equation into a quasiparticle contribution and
 incoherent background, we find:

\begin{equation}
G({\bf{k}},\omega) = \frac{Z_{\bf{k}}U_{\bf{k}}^{2}}{\omega - \Omega_{\bf{k}}
+i\delta} - 
\frac{Z_{\bf{k}}V_{\bf{k}}^{2}}{\omega + \Omega_{\bf{k}}-i\delta} + G_{inc}. 
\end{equation}
The renormalized triplet spectrum  and the renomalization constant are:
\begin{equation}
\Omega_{\bf{k}} = Z_{\bf{k}} \sqrt{( A_{\bf{k}} + \Sigma({\bf{k}},0))^{2}
- B_{\bf{k}}^{2}},
\end{equation}
\begin{equation}
Z_{\bf{k}}^{-1} = 1 - \left(\frac{\partial \Sigma}{\partial\omega}
\right)_{\omega =0} .
\end{equation}
The renomalized Bogoliubov coefficients in (11) are:
\begin{equation}
U_{\bf{k}}^{2},V_{\bf{k}}^{2} = \pm \frac{1}{2} +
\frac{ Z_{\bf{k}}( A_{\bf{k}} +\Sigma({\bf{k}},0))}{2\Omega_{\bf{k}}}.
\end{equation}
Equations (8,9,12-14) have to be solved self-consistently
for $\Sigma({\bf{k}},0))$ and $Z_{\bf{k}}$. From Eq.(11) it is also
clear that one has to replace $u_{\bf{k}} \rightarrow \sqrt{Z_{\bf{k}}}
U_{\bf{k}}, v_{\bf{k}} \rightarrow \sqrt{Z_{\bf{k}}} V_{\bf{k}}$
in (8) and (9) (and also in (15,16), see below).

We have found that effect of $H_{3}$ and $H_{4}$ on the quasiparticle 
spectrum is small, compared to the renormalization due
to $H_{U}$. However, these two terms have to be included for
the precise determination of the critical point.
We treat $H_{4}$ in mean field theory, by splitting the quartic
operator products into all possible pairs. This is equivalent to
taking only one-loop diagrams (first order in $J$) into account.
 These diagrams renormalize
the two coefficients:
\begin{equation}
A_{\bf{k}} \rightarrow A_{\bf{k}} +(1+\lambda) J\xi_{\bf{k}} \frac{1}{N}\sum_{\bf{q}}
\xi_{\bf{q}}v_{\bf{q}}^{2}, 
\end{equation}
\begin{equation}
B_{\bf{k}} \rightarrow B_{\bf{k}} -(1+\lambda) J\xi_{\bf{k}} \frac{1}{N}\sum_{\bf{q}}
\xi_{\bf{q}}u_{\bf{q}}v_{\bf{q}}.
\end{equation}
This concludes the solution of the two-layer problem $(\lambda = 1)$.

To solve the case  $\lambda = 0$ we also have to take into account    
$H_{3}$.
It is convenient to rewrite $H_{3}$ in terms of the 
Bogoliubov transformed operators $\tilde{t}_{{\bf{k}}\alpha},
\tilde{t}_{{\bf{k}}\alpha}^{\dagger}$, since in this way
only the  normal Green's functions remain.
To one loop order ($J^{2}$) the renormalization of the spectrum
is determined by the sum of the two diagrams in Fig.1c.
The formula for the interaction vertex in Fig.1c.  is quite lengthy  and 
we do not present it here.  Once the vertex is known, the self-energy
of Fig.1c. can be easily computed, leading to renormalization of
$A_{\bf{k}}$ and $B_{\bf{k}}$.

The results of the self-consistent numerical solution  are
summarized in Figures 2 and 3.
Figure 2 shows the triplet gap
$\Delta = \Omega_{\pi,\pi}$ as a function of  the interlayer coupling.
 The transition into the N\'{e}el ordered phase occurs at
$(J_{\perp}/J)_{c} = 2.57(\lambda = 1), 1.37(\lambda = 0)$.
We have also calculated the gap by using dimer series expansions 
\cite{Zheng}  up to  order 11(10) for $\lambda = 1$($\lambda = 0$).
 The critical points are found at: 
 $(J_{\perp}/J)_{c} = 2.52(2) (\lambda = 1),  1.39(4) (\lambda = 0)$,
or $2.537(5) (\lambda = 1), 1.393(8) (\lambda = 0)$ by
fixing the critical exponent $\nu=0.71$ \cite{Zheng}.
The agreement between the analytic method  and the dimer series  results is excellent. 
Such a  good agreement is better than might have been expected. 
Our analytic method 
involves approximations  and an error of a few percent
is always expected.  
 The gap critical exponent $\nu$, defined as:
$\Delta \sim (J - J_{c})^{\nu}$  is $\nu=0.5$ in our analytical calculation,
while the dimer series gives $\nu \approx 0.7$, in agreement with
the $O(3)$ non-linear sigma model prediction. Recall that  the  mean field
approximation gives $\nu=1$.
Our diagrammatic approach is not
valid very close to the  critical  point
since the neglected terms in $\Gamma(\bf{K})$ are of the
form 
 $\sum_{\bf q} v^2_{\bf q}/\omega_{\bf q}$ and
thus logarithmically diverge at criticality.
 However this  happens only very near to 
 the critical point.

The comparison of the excitation spectra, presented in  Figure 3,
shows that the agreement is very good over   almost  the whole
Brillouin zone. The disagreement between the two curves is  largest
at ${\bf{k}}=0$, where it is about 5\%. 

In conclusion, we have presented an effective analytical approach to take
into account the hard core constraint which appears in the  bond operator
description of the dimer phase. The triplet excitations are described as
a dilute Bose gas with infinite on-site repulsion.   
 We find that the spectrum is renormalized
mostly due to the  hard core, while the additional three and four point
interactions are comparatively weak and can be  treated perturbatively.
The advantages of our formulation are   that it is simple
 and captures the  essential physics, being in agreement within
a few percent with results obtained by dimer series expansions.
Obvious other applications of the method include  the
2D Heisenberg model with dimerization, the Heisenberg ladder and
the Kondo lattice model \cite{us}.
The method can also  be easily generalized 
  to describe phases with spontaneously broken symmetries
and nonzero temperature.

We wish to thank M. Kuchiev and R. Eder for stimulating discussions.
The financial support of the Australian Research Council is gratefully
acknowledged.

\vspace{9.0cm}

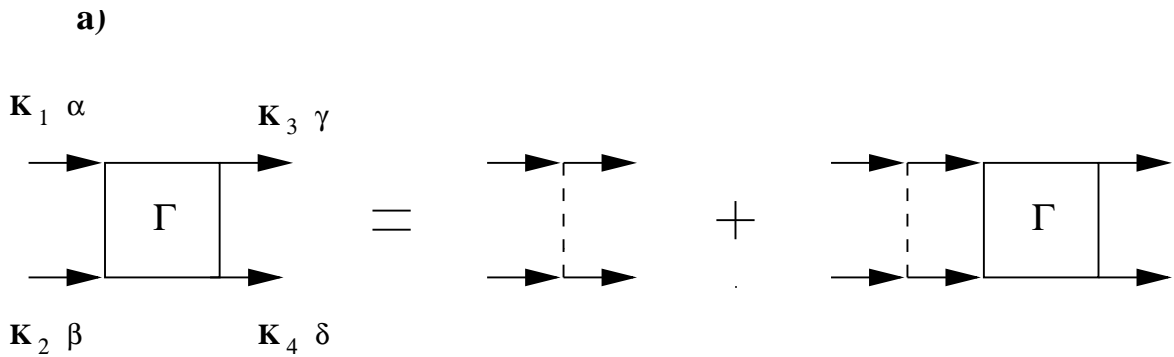
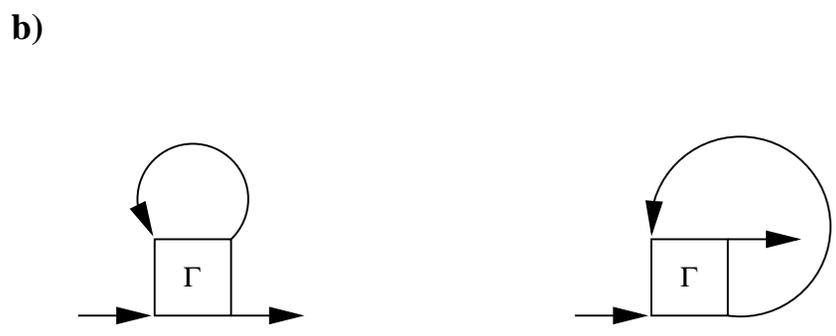
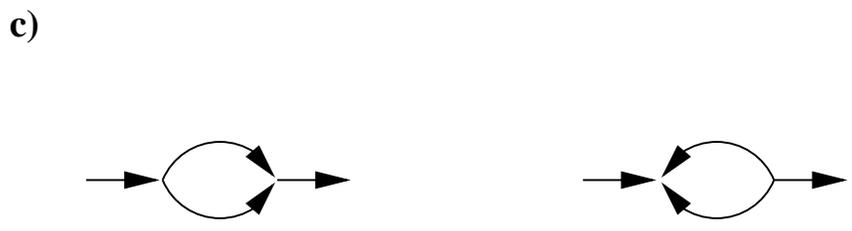
\begin{figure}
\caption
{(a) Equation for the scattering amplitude  $\Gamma$.
 (b) Diagrams for the self-energy, corresponding to $\Gamma$.
 (c) One loop diagrams, arising from the three-point interaction.  }
\label{fig.1}
\end{figure}

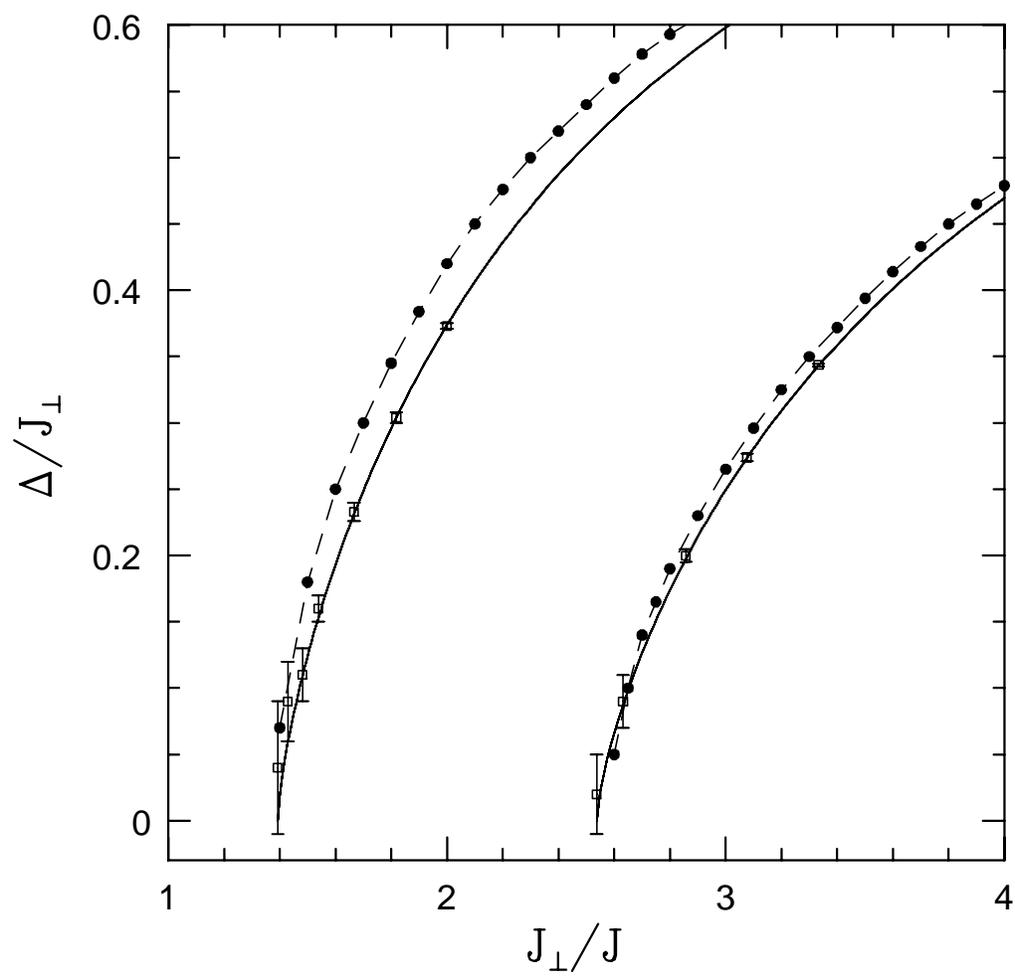
\begin{figure}
\caption
{Triplet gap as a function of interlayer coupling for $\lambda =0$
(left curves) and $\lambda =1$ (right curves).
 The dashed lines with
the solid circles are the results of the self-consistent solution.
Open squares (with error bars) are from direct Pade approximants
to the dimer series while solid lines are from approximants which assume
$\nu = 0.71$.  }
\label{fig.2}
\end{figure}
 
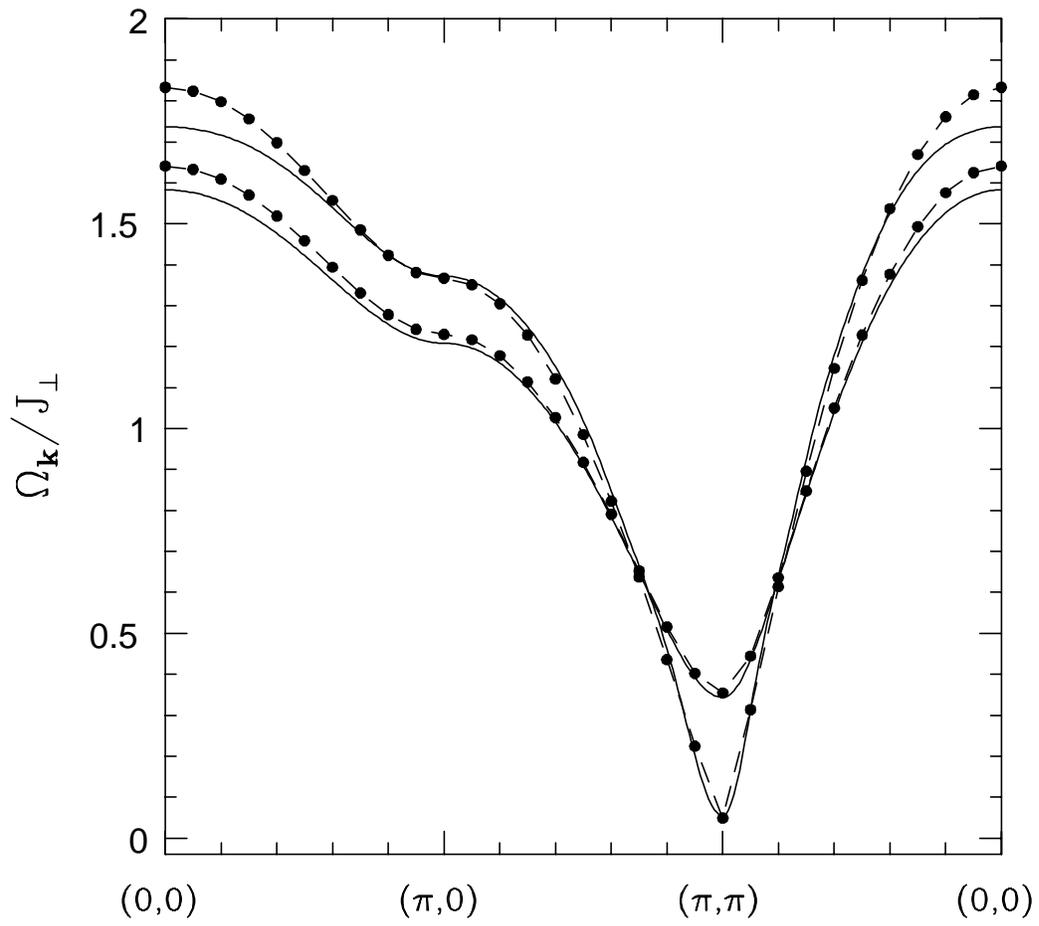
\begin{figure}
\caption
{Triplet excitation spectrum for $\lambda =1$ 
 along high symmetry directions  in the Brillouin zone.
The dashed line with solid circles is the self-consistent solution while the
 solid lines are from direct summation of the dimer series.
 The upper (at ${\bf{k}}=0$)
 dimer series curve
corresponds to the critical (within the error bar) spectrum ($J_{\perp}/J = 2.54$),
while the upper analytical curve is computed at $J_{\perp}/J =2.6$, in order to
have the same gap $\Delta/J_{\perp} = 0.05$. The lower (at ${\bf{k}}=0$)
 curves correspond
to  $J_{\perp}/J = 3.33$.       
}
\label{fig.3}
\end{figure}

\end{document}